\documentclass[conference]{IEEEtran}
\def\BibTeX{{\rm B\kern-.05em{\sc i\kern-.025em b}\kern-.08em
    T\kern-.1667em\lower.7ex\hbox{E}\kern-.125emX}}

\usepackage{ifpdf}

\usepackage{tikz}
\usepackage{textcomp}
\usepackage{hyperref}
\usepackage{lipsum}

\usepackage{cite}
\usepackage[english]{babel}
\usepackage[utf8]{inputenc}

\usepackage[nottoc]{tocbibind}

\usepackage{color}

\usepackage{dirtytalk}
\usepackage{graphicx}
\usepackage{amsmath}
\usepackage{amssymb}

\usepackage{algorithmic}

\usepackage{array}

\ifCLASSOPTIONcompsoc
  \usepackage[caption=false,font=normalsize,labelfont=sf,textfont=sf]{subfig}
\else
 \usepackage[caption=false,font=footnotesize]{subfig}
\fi

\usepackage{stfloats}
\begin{document}
%

\title{Power Flow Approximations for Multiphase Distribution Networks using Gaussian Processes}


\author{Daniel Glover,~\IEEEmembership{Graduate Student Member, IEEE},
Parikshit Pareek,~\IEEEmembership{Member, IEEE},\\
Deepjyoti Deka,~\IEEEmembership{Senior Member, IEEE},
Anamika Dubey,~\IEEEmembership{Senior Member, IEEE}
}


%



\IEEEoverridecommandlockouts

\newcommand\copyrighttext{
\footnotesize IEEE Copyright Notice. \textcopyright IEEE 2025. This paper is accepted for publication in IEEE PES General Meeting 2025. The complete copyright version will be available in IEEE Xplore with published conference proceedings. Personal use of this material is permitted.
Permission from IEEE must be obtained for all other uses, in any current or future
media, including reprinting/republishing this material for advertising or promotional purposes, creating new collective works, for resale or redistribution to servers or lists, or reuse of any copyrighted component of this work in other works.}

\newcommand\copyrightnotice{%
\begin{tikzpicture}[remember picture,overlay]
\node[anchor=south,yshift=10pt] at (current page.south) {\fbox{\parbox{\dimexpr\textwidth-\fboxsep-\fboxrule\relax}{\copyrighttext}}};
\end{tikzpicture}%
}
\maketitle
\copyrightnotice

\begin{abstract}
Learning-based approaches are increasingly leveraged to manage and coordinate the operation of grid-edge resources in active power distribution networks. Among these, model-based techniques stand out for their superior data efficiency and robustness compared to model-free methods. However, effective model learning requires a learning-based approximator for the underlying power flow model. This study extends existing work by introducing a data-driven power flow method based on Gaussian Processes (GPs) to approximate the multiphase power flow model, by mapping net load injections to nodal voltages. Simulation results using the IEEE 123-bus and 8500-node distribution test feeders demonstrate that the trained GP model can reliably predict the nonlinear power flow solutions with minimal training data. We also conduct a comparative analysis of the training efficiency and testing performance of the proposed GP-based power flow approximator against a deep neural network-based approximator, highlighting the advantages of our data-efficient approach. Results over realistic operating conditions show that despite an $85\%$ reduction in the training sample size (corresponding to a $92.8\%$ improvement in training time), GP models produce a $99.9\%$ relative reduction in mean absolute error compared to the baselines of deep neural networks.
\end{abstract}

\section{Introduction}

Traditional power distribution systems are rapidly expanding into active distribution networks (ADNs), operating as massive cyber-physical systems managing both bidirectional energy delivery and information traffic.  Heavy penetrations of stochastic distributed energy resources (DERs) and unpredictable load variations, combined with partial observability at the grid-edge, have created complexities for operators in coordinating these resources. 
Utility companies and operators must proactively coordinate distribution-connected controllable resources to ensure efficient and reliable grid operations, especially with the integration of DERs at the grid edge.  State-of-the art methods formulate an optimal power flow (OPF) problem to coordinate numerous grid-edge active nodes. The nonlinear power flow (PF) equations render the resulting OPF problem a nonlinear constrained optimization problem which is computationally expensive \cite{dubey2023distribution, molzahn2017computing}. To simplify OPF problems, approximate power flow models have been proposed for unbalanced multiphase ADNs \cite{deka2019topology, gan2014convex}.  However, linearized models may erroneously approximate bus voltages, especially during high loading conditions \cite{inaolaji2021}.

The computational challenges of solving distribution OPF problems has motivated the use of learning-based approaches to control and optimize the active nodes in power distribution systems \cite{yang2023optimal}. In this context, both model-free and model-based learning-based approaches have been proposed to optimize distribution system operations.  Model-free approaches do not employ any analytical model of power system physics when learning to control or optimize grid operations; they directly learn control policy from data. On the contrary, model-based approaches, while learning the control policy, also simultaneously learn a model of the underlying power systems physics. Model learning entails using different supervised learning techniques to learn power flow solutions (voltage and angle) given a set of active and reactive power injections. The learned models serve as a learning-based approximator for the power flow problem. Generally, these learning-based power flow approximators are based on deep neural networks (DNNs) which are data hungry. This work presents a data-efficient approach for learning-based approximation of distribution PF utilizing Gaussian Processes.

Alternative approaches for PF approximation from the machine learning community have shown promise through the use of neural networks (NNs) \cite{khodayar2020deep, tiwari2024power}. Authors in \cite{hu2020physics} use an encoder/decoder with regularization to build a physics-guided DNN PF solver with tractability of parameter knowledge.  A probabilistic PF model is designed in \cite{yang2019fast} incorporating the branch flow equations into the loss function of a DNN to improve solution feasibility, and \cite{hansen2022power} utilizes graph neural networks (GNNs) representing distribution feeders as branch graphs to locally estimate power balance in the face of topology shifting. Although deep learning approaches can achieve high performance, they come at the cost of constant retraining and large data demand, which may not be available to the utility.  DNNs also remain less interpretable to system operators, and are typically viewed as inherently \textit{black-box}.

This work focuses on Bayesian methods, which differ from DNNs in that they encode a conditional probability distribution over the data, providing a measure of uncertainty against new predictions. Bayesian methods have been used previously for different analytical tasks in ADNs. \cite{xu2021adaptive} proposes an importance sampling scheme under the Bayesian framework to simultaneously estimate the topology, outages, and power injections of a distribution system.  A probabilistic PF model from \cite{zuluaga2018bayesian} proposes a prior distribution over state variables and influences PF solutions using a derived Jacobian matrix to improve the posterior distribution, outperforming Monte Carlo (MC) methods. Similarly, \cite{pareek2020gaussian} develops a sensitivity analysis for probabilistic OPF using Gaussian Processes (GPs) to overcome uncertain renewable injections, and \cite{pareek2021framework} proposes a novel closed-form power flow for single-phase systems under unpredictable load dynamics using GPs to explain voltage fluctuation influence. Overall, Bayesian frameworks offer flexibility and transparency when compared to NNs, particularly when training data size is manageable.  

In this paper, we extend a GP PF approximator to three-phase unbalanced power distribution systems to provide insight into real-world grid operational challenges.  We conduct a comparative case study between GPs and DNNs that is beneficial to utilities looking to adopt learning-based grid-edge control.  We test the accuracy of the proposed GP-based PF for realistic testing scenarios and for different amounts of training data on the IEEE 123-bus test system with multiple DERs and variable loads.  Furthermore, the scalability of the trained GP PF approximator utilizing minimal training data is evaluated on the IEEE 8500-node system. Given the prevalence of using DNN-based approaches for PF, we also present an extensive comparative study on training efficiency and testing performance for GP-based and DNN-based power flow solvers. Both learning-based models are compared against a linearized power flow and a nonlinear power flow model.  Our results demonstrate that GPs are superior in terms of training efficiency and accuracy under limited measurements compared to DNN PF models.
\vspace{-1.25mm}

\section{Gaussian Process Regression \& Power Flow}
\subsection{Three-Phase Distribution System Model}
The distribution system is modeled as a directed graph $\mathcal{G}=(\mathcal{N},\mathcal{E})$, where $\mathcal{N}$ and $\mathcal{E}$ denote the set of nodes (buses) and edges (lines). Two nodes $i$ and $j$ are connected to form an edge $(i,j)$ whenever node $i$ is the parent node for node $j$. For a node $i$ in the system, the associated three-phase equivalent is denoted by $\rho_{i}, \phi_{i} \in \lbrace a,b,c \rbrace $. At node $i \in \mathcal{N}$ for a phase $\rho$, the complex voltage phasor is denoted by $V_i^{\rho} \angle \theta_i^{\rho}$, where $V_i^{\rho}$ is the magnitude and $\theta_i^{\rho}$ the angle, with voltage vector $\bf{V}$ and angle vector $\boldsymbol{\theta}$.  The apparent power of the load is defined by $S_{Li}^{\rho}=P_{Li}^{\rho}+j Q_{Li}^{\rho}$, and the net load at each node per phase $s_{Li}^{\rho} \in \mathbf{s}$ is equivalent to the real and reactive power of the load and the negative of local DER injections $S_{L,i}^{\rho} - S_{DER,i}^{\rho} $, represented by the joint column vector $\bf{s}=\left[\bf{p};\bf{q} \right]$.  Using the functional representation from \cite{pareek2021framework}, we represent the family of nonlinear power flow equations as the mapping between the vector of PF solutions $\bf{z}=\left[\bf{V};\boldsymbol{\theta}\right]$ and net loads $\textbf{z}=F(\textbf{s})$.  The nonlinear equations $F(\cdot)$ are generally solved iteratively via Newton-Raphson or Backward-Forward Sweep, however, the goal here is to use GP regression to obtain an approximate PF solution to $F(\textbf{s})$, mapping the relationship between the load vector $\textbf{s}$ and the voltage magnitude solution $\mathbf{V}$ (or $\boldsymbol{\theta}$).

\subsection{Gaussian Process Regression}
A Gaussian process (GP) is a nonparametric regression method which describes a prior probability distribution over the space of possible functions $f_i(x)$ that fits data set $\mathcal{D}=\{(x_i, {y}_i)|i=1,...,n\}$, with input data $X=[x_1,...,x_n]$ and targets $\mathbf{y}=[y_1,...,y_n]$ \cite{williams2006gaussian}. The GP is defined by a \textit{mean} $m(x) = \mathbb{E}[f(x)]$ and \textit{covariance function} $k(x^i, x^j)$, where $f(x^i)$ and $f(x^j)$ are jointly Gaussian, given as $k(x^i,x^j) = \mathbb{E}\bigl[ (f(x^i) - m(x^i))(f(x^j) - m(x^j)) \bigr]$. Then the GP for the function $f(x)$ of random variables is given as:
\begin{equation}
f(x) \sim \mathcal{GP}(m(x), k(X,X))
\end{equation}
The covariance function, or \textit{kernel}, defines the similarity between two inputs as a measure of variance and dictates the \say{smoothness} of the function fit to $\mathcal{D}$.  The \textit{squared exponential kernel} (radial basis function or RBF kernel) used in this study is given in (2), where the hyperparameters to be optimized are $l$ \textit{lengthscale} and signal \textit{variance} $\sigma_s^2$. 
\begin{equation}
k(x^i,x^j) = \sigma_s^2 \exp \Big [\frac{-\|x^i - x^j\|^2}{2l} \Big ]
\end{equation}
The target data $\mathbf{y}$ is considered to be GP values corrupted with i.i.d. Gaussian noise $\epsilon \sim \mathcal{N}(\mu=0,\,\sigma_{\epsilon}^{2})$. The GP thus assumes a Gaussian distribution for observed data $\mathcal{D}$ in the form of $y \sim \mathcal{N}(m(X),k(X,X)+ \sigma^2_\epsilon\mathbb{I})$, the marginal log likelihood based optimized hyperparameters are given by $\Theta^*(\sigma_s^2, l) = arg \max_{\Theta} \log pr(\mathbf{y}|X,\Theta)$.  For a new input $x^*$, the mean multiphase PF prediction $\hat{f}$ at $x^*$ is given as 
\begin{equation}
\hat{f}(x^*)|X,\mathbf{y},x^*,\Theta^* \sim \mathcal{N}(\hat{f}(x^*),cov(f^*(x^*)))
\end{equation}
where $cov(f^*)$ comes from the joint probability distribution of the observed values and the mean function values at new testing points.  Under the Gaussianity assumption, the GP prediction given a new data point $x^*$ takes the analytical form
\begin{equation}
\hat{f}(x^*) = \mathbf{k}^T (K + \sigma_{\epsilon}^2 I)^{-1} \mathbf{y}
\end{equation}
where, $\mathbf{k} = k(X,x^*)$ is the kernel vector mapping the new data point variance to all inputs, and $I$ is the identity matrix.  The continuous and differential properties of GPs make them an excellent tool to approximate injection to voltage maps.


\section{Methodology \& Learning Models}
\subsection{Gaussian Process Power Flow Approximation Model}
In our model, the GP provides a PF solution prediction of $\mathbf{y}=\mathbf{V}$ given a new set of net load injection vectors $\mathbf{s}^*$ across multiple phases concurrently.  To train the GP, we gather net load data per node per phase in the \textit{design matrix} $X = [s_{Li}^{\rho},...,s_{Ln}^{\rho}] = [p_{Li}^{\rho},q_{Li}^{\rho},...,p_{Ln}^{\rho},q_{Ln}^{\rho}]$ and the corresponding nonlinear PF solution per phase of voltage magnitudes $\mathbf{y}=[V_i^{\rho},...,V_n^{\rho}]$ (or angles $\theta_i^{\rho}$).  The input vector of nodal net load quantities is represented as the real and reactive net loads per phase. 

\subsection{Benchmark PF Models: DNN and LinDistFlow}
To compare supervised learning approaches, a fully connected DNN is constructed using \textit{multi-layer perceptron} hidden layers to approximate the multiphase nonlinear PF, shown in Figure 1.  The DNN takes as input $\mathbf{s}^T$ and outputs a PF prediction $\mathbf{y}^T$, parameterizing the weights of each layer $\boldsymbol{\theta}_i$ with each backward pass over the training set $\mathcal{D}$. The \textit{backpropagation} procedure utilizes a traditional \textit{mean squared error} (MSE) loss function $\mathcal{L}_{MSE} \leftarrow \frac{1}{n} \sum_{i=1}^{n} \left(\hat{y}_i - y_i\right)^2$  to update $f(\boldsymbol{\theta})$ in the direction of the gradient.  A summary of model parameters is given in Tables I and II.

\begin{figure}[!t]
    \centering
    \includegraphics[width=2.5in]{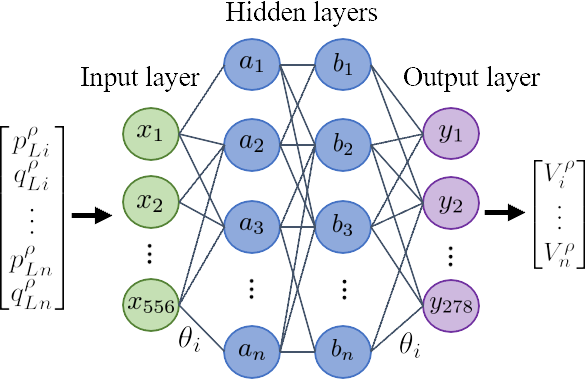}
    \caption{Deep Neural Network Architecture}
    \label{fig:dnn}
    \vspace{-4mm}
\end{figure}

The linear three-phase distribution power flow (LDF) model \textit{LinDistFlow} is presented for reference as a means of standardized benchmarking against the GP and DNN.  The LDF yields active and reactive power balance equations presented in (5) and (6), where $k: j \rightarrow k$ denotes a child node $k$ for a parent node $j$, and (7) represents a linearized three-phase voltage drop equation, subject to $(v_{min})^2 \leq v_{i,t}^{\rho} \leq (v_{max})^2$.  For an edge $(i,j) \in \mathcal{E}$ with phase $\rho$, the apparent power flow and complex current flow are shown by $S_{ij}^{\rho}$ and $I_{ij}^{\rho}$ with line impedance $z_{ij}^{\rho}$, respectively.  The state variables are active power flow, $P_{ij}^{\rho\rho}$, reactive power flow, $Q_{ij}^{\rho\rho}$, and voltage magnitude squared, $v_{i}^{\rho}$. 
\vspace{-3mm}
\begin{eqnarray}
    P_{ij}^{\rho \rho} - p_{L,j}^{\rho} &=& \sum_{k; j \rightarrow k} P_{jk}^{\rho \rho} \hspace{0.5 cm}  \rho \in \lbrace a,b,c \rbrace\\    
    Q_{ij}^{\rho \rho} - q_{L,j}^{\rho} &=& \sum_{k; j \rightarrow k} Q_{jk}^{\rho \rho}  \hspace{0.5 cm}  \rho \in  \lbrace a,b,c \rbrace \\
    v_{i}^{\rho} - v_{j}^{\rho} &=& \sum_{q \in \phi_{j}} 2\mathbb{R}[S_{ij}^{pq} (z_{ij}^{pq})^*]     p,q \in  \lbrace a,b,c \rbrace
\end{eqnarray}

\section{Simulation Case Study}
\subsection{Distribution Simulation Environment}

To perform data acquisition and comparatively evaluate the models in a realistic distribution system, we utilize OpenDSS \cite{dugan2011open} to construct the modified 123-bus test system in Figure 2, simulating a quasi-static time series. The IEEE 123-bus system is selected for this study due to its standardized unbalanced three-phase representation of a real-world radial distribution feeder with multiple DERs.  In total, there are 99 phase A nodes, 84 phase B nodes, and 95 phase C nodes, yielding an input vector $\bf{s}$ of shape $[556,]$ and an output vector $\bf{y}$ of shape $[278,]$. The feeder contains twenty solar PV systems (both single and three-phase) distributed at random load locations and sized according to local hosting capacity demand.

\begin{figure}[!t]
    \centering
    \includegraphics[width=3.45in]{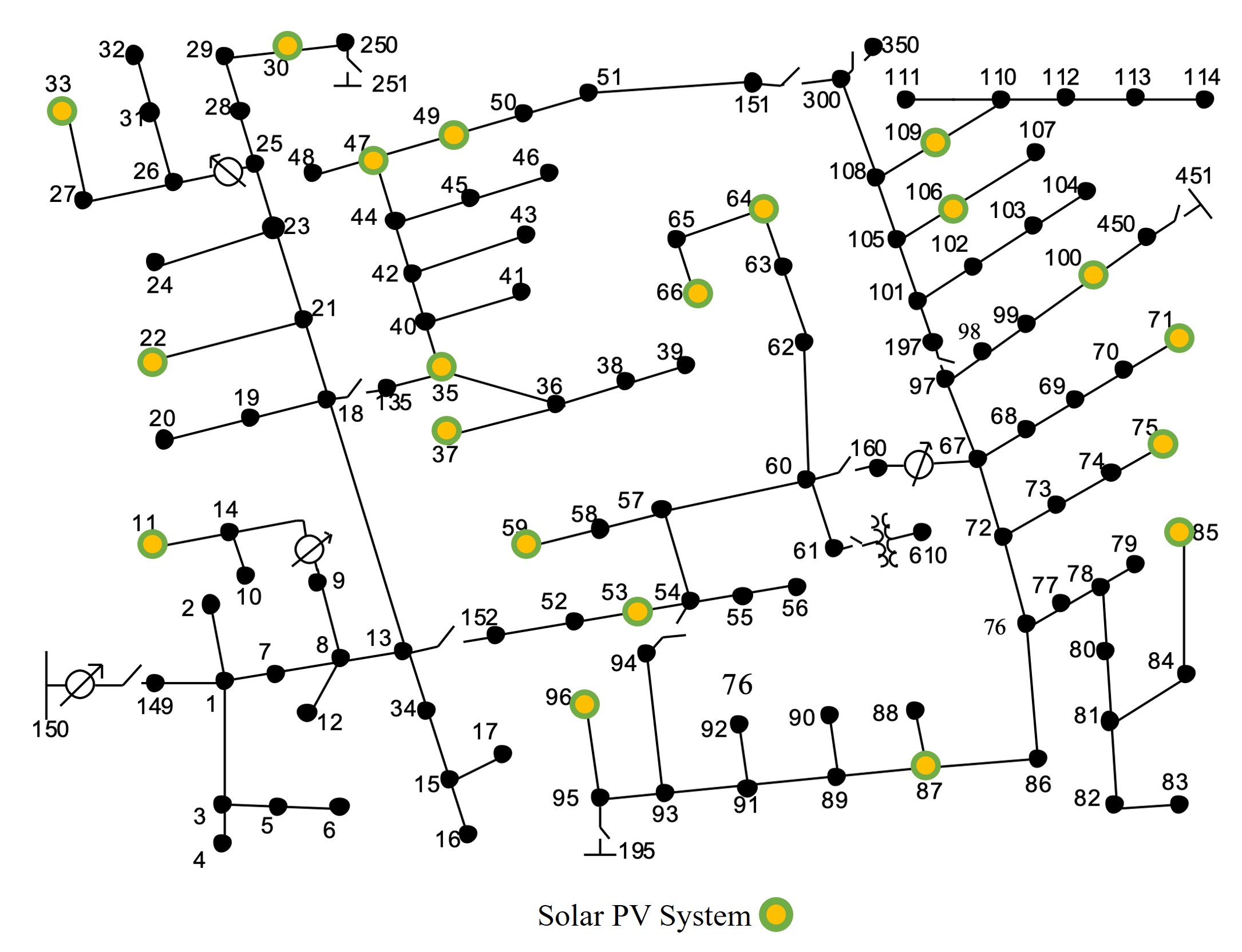}
    \caption{IEEE 123-Bus Test System with DERs}
    \label{fig:123bus}
    \vspace{-2mm}
\end{figure}

\begin{table}[!t]
    \centering
    \caption{Gaussian Process Regression Model Parameters}
    \label{tab:GPParameters}
    \begin{tabular}{c|c|c|c|c}    
        \hline
        \bf{Model}  & \multicolumn{4}{c}{\bf{MIMO GP}}   \\
        \cline{2-5}
       \bf{Parameters} & Case 1 & Case 2 & Case 3 & Case 4  \\
       \hline
        \cline{2-4}
        Kernel & RBF & RBF & RBF & RBF \\
        \hline
        Optimizer  & L-BFGS-B  & L-BFGS-B  & L-BFGS-B  & L-BFGS-B \\
        \hline
        ARD & dims & dims & dims & dims \\
        \hline
        GNV & True & True & True & True \\
        \hline
        $l$, $\sigma_s^2$ & 1 & 1 & 1 & 1\\
        \hline
    \end{tabular}
   \vspace{-2mm}
\end{table}
\vspace{-2mm}
\begin{table}[!t]
    \centering
    \caption{Deep Neural Network Hyperparameters}
    \label{tab:DNNParameters}
    \begin{tabular}{c|c|c|c|c}    
        \hline
        \bf{Hyper}  & \multicolumn{4}{c}{\bf{Fully Connected DNN}}   \\
        \cline{2-5}
       \bf{Parameters} & Case 1 & Case 2 & Case 3 & Case 4  \\
       \hline
        \cline{2-4}
        Input Layer & [556,] & [556,] & [556,] & [556,] \\
        \hline
        Hidden Layer 1 & [556,556] & [556,556] & [556,556] & [556,556] \\
        \hline
        Hidden Layer 2 & [556,278] & [556,278] & [556,278] & [556,278] \\
        \hline
        Output Layer & [278,] & [278,] & [278,] & [278,] \\
        \hline 
        Layer Type & \textit{MLP} & \textit{MLP} & \textit{MLP} & \textit{MLP} \\
        \hline
        Optimizer  & Adam  & Adam  & Adam  & Adam \\
        \hline
        Regularization & Yes & Yes & Yes & Yes \\
        \hline 
        Activation Function & GELU & GELU & GELU & GELU \\
        \hline
        Learning Rate & 0.0001 & 0.0001 & 0.0001 & 0.0001 \\
        \hline
        Batch Size & 32 & 32 & 32 & 32  \\
        \hline
        Epochs & 400 & 400 & 400 & 400 \\
        \hline
        Parameter Noise & 0.05 & 0.05 & 0.05 & 0.05 \\
        \hline
    \end{tabular}
    \vspace{-2mm}
\end{table}

\vspace{2mm}
We consider a realistic case study based on acquired hourly net load injection data quantities available to the utility in four separate durations: 1 day (24 hrs), 7 days (168 hrs), 30 days (720 hrs), and 90 days (2160 hrs).  In each case, the GP and DNN are trained on the available data and evaluated against the nonlinear OpenDSS PF solution and LDF model, noting both the \textit{mean squared error} (MSE) $\mathcal{L}_{MSE} \leftarrow \frac{1}{n} \sum_{i=1}^{n} \left(\hat{y}_i - y_i\right)^2$ and \textit{mean absolute error} (MAE) $\mathcal{L}_{MAE} \leftarrow \frac{1}{n} \sum_{i=1}^{n} |\hat{y}_i - y_i|$  solution accuracy, as well as the compute time required to train the models.  Three different loadshape types (residential, commercial, and industrial) are applied randomly over all loads, and a variability parameter with load multiplier is used to fluctuate overloading at each time step following a PF solution.  Real solar irradiance and temperature data from central Texas, USA is applied to all DERs from \cite{sengupta2018national}, and the default regulator and capacitor bank controls are disengaged to allow for voltage fluctuations due to loading and DER injections.  All training is performed on an Intel(R) Core i7-7500 CPU with 16GB RAM.    


\begin{table*}[!t]
    \centering
    \caption{Case Study Performance Metrics 123-Bus: Mean Squared Error, Mean Absolute Error \& Computational Training Time}
    \label{tab:Parameters}
    \begin{tabular}{|c|c|c|c|c|c|c|c|c|c|c|c|c|}    
        \hline
        \bf{PF} & \multicolumn{3}{|c|}{\bf{Case 1: 1 Day, 24 hrs}} & \multicolumn{3}{|c|}{\bf{Case 2: 7 Days, 168 hrs}} & \multicolumn{3}{|c|}{\bf{Case 3: 30 Days, 720 hrs}} & \multicolumn{3}{|c|}{\bf{Case 4: 90 Days, 2160 hrs}}    \\
        \cline{2-13}
       \bf{Est.} & MSE & MAE & Time & MSE & MAE & Time & MSE & MAE & Time & MSE & MAE & Time\\
        \hline
        \bf{LDF} & 0.0013 & 0.053 & 00:00:20 & 0.00013 & 0.0069 & 00:03:30 & 3.1e-05 & 0.0016 &  00:15:02 & 1.1e-05 & 5.6e-04 & 00:55:44 \\ 
        \hline
        \bf{DNN}  & 0.02 & 0.1197 & 00:00:14 & 0.0103 & 0.0514 & 00:00:25& 7.14e-06 & 0.0023 & \bf{00:00:43} & 4.12e-08 & 1.48e-04 & \bf{00:01:51}   \\
        \hline
        \bf{GP}  & \bf{1.67e-09} & \bf{2.9e-05} & \bf{00:00:01} & \bf{3.94e-10} & \bf{1.3e-05} & \bf{00:00:04} & \bf{6.53e-11} & \bf{5.44e-06} & 00:00:44  & \bf{2.22e-11} & \bf{2.93e-06} & 00:27:21  \\
        \hline
        \end{tabular}
   \vspace{-2mm}
\end{table*}

\begin{figure}[!t]
    \vspace{-2mm}
    \centering
    \includegraphics[width=3.45in, height=2.0in]{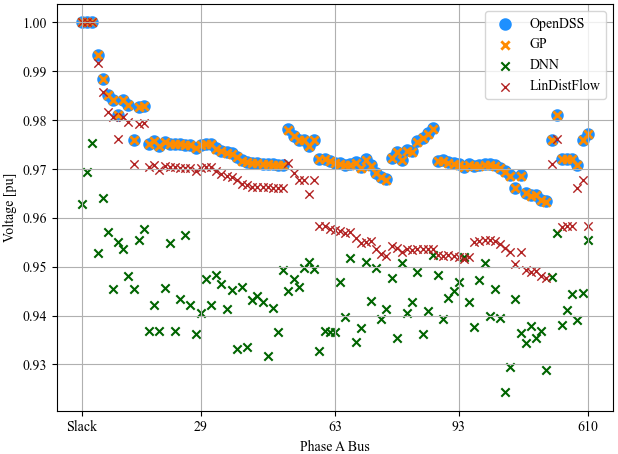}
    \caption{Case 1: Phase A Voltage Magnitude Prediction Hour 33}
    \label{fig:case1_voltages}
    \vspace{-2mm}
\end{figure}

\begin{figure}[!t]
    \vspace{-2mm}
    \centering
    \includegraphics[width=3.45in, height=2.0in]{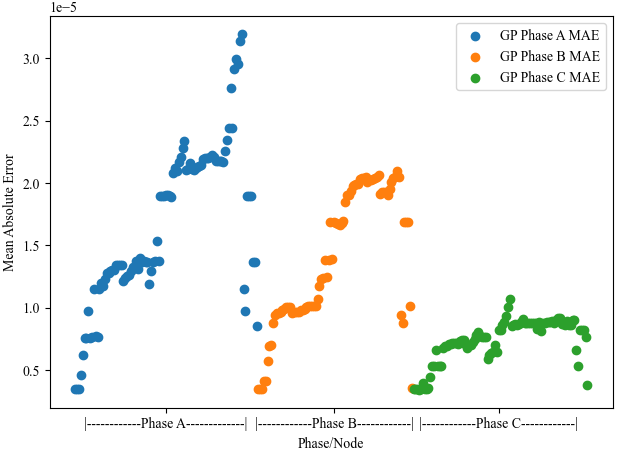}
    \caption{Case 2 Training: GP Mean Absolute Error Three Phases}
    \vspace{-1mm}
    \label{fig:case2_gpmae}
    \vspace{-2mm}
\end{figure}

\begin{figure}[!ht]
    \vspace{-1mm}
    \centering
    \includegraphics[width=3.45in, height=2.0in]{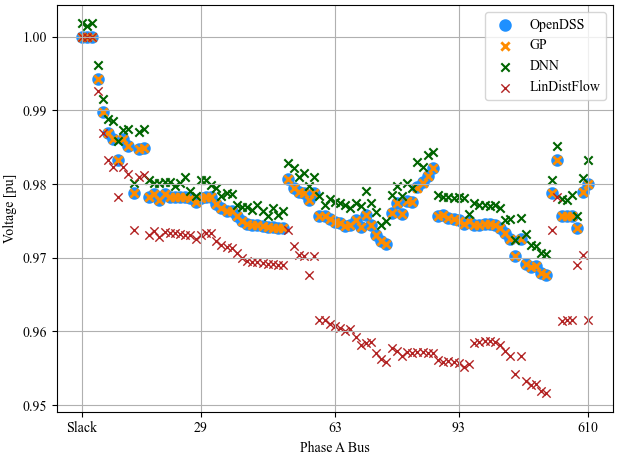}
    \caption{Case 3: Phase A Voltage Magnitude Prediction Hour 107}
    \vspace{-1mm}
    \label{fig:case3_voltages}
    \vspace{-2mm}
\end{figure}

\begin{figure}[!ht]
\vspace{-1mm}
    \centering
    \includegraphics[width=3.45in, height=2.0in]{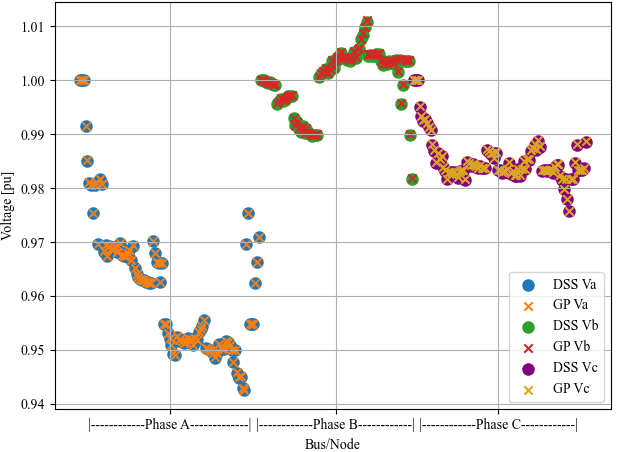}
    \caption{24-Hour GP Model Test Day 5, Hour 112: $3\phi$ Voltage Prediction}
    \vspace{-2mm}
    \label{fig:case1_gp_3phvolts}
    \vspace{-2mm}
\end{figure}

\subsection{Model Evaluation \& Discussion}
Case study results from training and validation on the 123-bus system are summarized in Table III with training times listed in hrs:min:sec.  Figures 3 and 5 show two, A-phase GP hourly PF snapshot testing predictions for clarity, with the OpenDSS bus label ID (x-axis) moving out from the substation versus power flow voltage magnitude (y-axis). In case 1, the utility has access only to a set of 24 measurement vectors, rendering the DNN useless (see Figure 3). The GP however, trains in a single second and obtains PF solution accuracy of no worse than 7.9e-04 when compared to the nonlinear solution. In case 2, the GP exhibits a prediction MSE of 3.94e-10 and MAE of 1.03e-05, again outperforming the DNN (and LDF).  MAE for predictions across all three phases for case 2 in Figure 4 shows the maximum error occurring at bus locations furthest from the substation, where the voltages are more variable. The training speed of the GP here is also an 84\% improvement over the DNN.

From Figure 5, case 3 displays significant improvement on the part of the DNN accuracy with more available data.  Although the GP still remains the least error prone, its training times have now dramatically increased.  In case 4, the utility now has access to a much larger training set of 2160 observation vectors, and both the GP and DNN show extremely high accuracy of the PF solution.  More importantly in this case, there exists a notable difference in computational cost required to train the models, with the DNN taking just under two minutes to train compared to the GP at nearly thirty minutes! Nonetheless, this concern is irrelevant from a practical grid operational standpoint because 1) the GP accuracy is still highest, and 2) the GP training time can be reduced simply by sampling the larger data set to achieve an acceptable PF prediction, as is represented in cases 1 and 2.

To further validate our method, a pre-trained GP using only 24 hours of data is tested over a consecutive six day span at every hour.  Results in Table IV confirm a maximum error of 5.95e-04 and a minimum of 7.9e-11, with low average MAE and MSE over all testing samples.  Figure 6 shows a single PF prediction at hour 112 of testing (day 5), demonstrating the robustness of the GP PF approximator across all three phases considering uncertain loading and DER penetrations.  Finally, we examine the scalability of the GP using the IEEE 8500-node system in a similar six-day testing scenario, predicting the PF solution across 3823 primary phases.  Table V shows the GP prediction average MAE and MSE on all testing points is again very small, and is likely due in this case to the absence of loading and DERs on the primary feeder preventing larger variances.  However, the GP still remains very effective in PF approximation considering the feeder size and limited data requirements.   Figure 7 shows all 1216 primary phase C buses and the GP PF testing predictions versus OpenDSS solutions.  These results are critical in that they solidify the GP method as a fast, reliable PF approximator that can scale to larger feeder laterals accurately with minimal data under uncertainty.  In the big picture, these improvements translate to more efficient real-world grid operations for utilities managing DER dominated ADNs.  With GPs, operators can reliably estimate the system state or solve the OPF under limited data, using a transparent learning-based model with minimal incurred cost.

\begin{table}[!t]
    \centering
    \caption{24-Hour GP Model, 123-Bus System: Six-Day Testing Period}
    \label{tab:GPtest_1Day}
    \begin{tabular}{|c|c|c|c|}    
        \hline
        \multicolumn{4}{|c|}{\bf{IEEE 123-Bus Trained GP Evaluation}}   \\
        \hline
          Avg MAE & Avg MSE & Maximum Error & Minimum Error  \\
       \hline
       $2.869e-04$ & $4.2555e-06$ & $0.000595$ & $7.91351e-11$\\
       \hline
    \end{tabular}
  \vspace{-2mm}
\end{table}

\begin{table}[!t]
    \centering
    \caption{24-Hour GP Model, 8500-Node System: Six-Day Testing Period}
    \label{tab:GPtest_1Day}
    \begin{tabular}{|c|c|c|c|}    
        \hline
        \multicolumn{4}{|c|}{\bf{IEEE 8500-Node Primary Feeder Trained GP Evaluation}}   \\
        \hline
          Avg MAE & Avg MSE & Maximum Error & Minimum Error  \\
       \hline
       $5.833e-05$ & $8.4722e-09$ & $0.000308$ & $1.4475e-12$\\
       \hline
    \end{tabular}
  \vspace{-2mm}
\end{table}
\begin{figure}[!t]
\vspace{-1mm}
    \centering
    \includegraphics[width=3.45in]{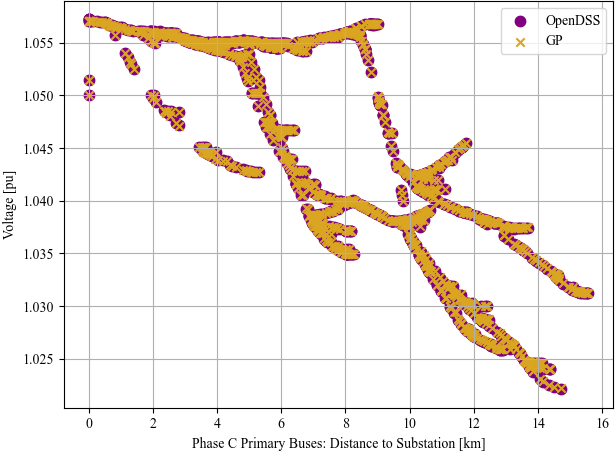}
    \caption{IEEE 8500-Node System Test Day 2, Hour 37: Phase C Voltage}
    \vspace{-2mm}
    \label{fig:case1_gp_phc_8500}
    \vspace{-2mm}
\end{figure}

\vspace{-1.1mm}
\section{Conclusion}
In conclusion, this study has shown the capabilities of utilizing GPs against deep learning models for learning the nonlinear PF solution in multiphase ADNs.  A comparative case study shows that GPs have the ability to map net injection information to the correct voltage magnitude PF solution in unbalanced multiphase networks subject to high penetrations of DERs and variable load fluctuations.  When data is limited to distribution operators, a GP is recommended to provide quick, accurate solutions at a fraction of the time taken to train a DNN, with much higher efficiency due to the high data demand requirements of DNNs. Additionally, the nonparametric nature of the GP requires no model retraining and can be maintained by operators to make new predictions on the fly. Although GPs do require adequate computing resources to train on larger datasets, this issue can be easily mitigated by batch sampling to improve compute times. This trade off, in our estimation, is more cost prohibitive to the utility when compared to the data acquisition and retraining requirements of DNNs.  Regarding future work direction, the authors aim to examine model scalability in larger feeders, explore alternative kernel designs, and incorporate topology reconfigurations into GP model training to better understand the impact of data reduction with respect to network topology shifts.

\vspace{-1.0mm}

%
\IEEEpeerreviewmaketitle





%

\bibliographystyle{IEEEtran}
\footnotesize{\bibliography{references}}






\end{document}